\documentstyle[12pt,fleqn]{article}
\newcommand{\spc}{\!\!\!\!\!\!\!\!\!\!\!}
\makeatletter
\@addtoreset{equation}{section}
\makeatother

\begin{document}
\begin{center}
\huge\bf Bohmian Mechanics Revisited \\
\vspace*{0.4in}
\normalsize\bf E.\  Deotto \\
\small \rm Department of Theoretical Physics, University of Trieste,
Italy. \\
\vspace*{0.1in}
and \\
\vspace*{0.1in}
\normalsize \bf G.\ C.\ Ghirardi\\
\small \rm Department of Theoretical Physics, University of Trieste,\\
International Centre for Theoretical Physics, Trieste, Italy.
\end{center}
\vspace{0.2in}
\bf Abstract:\
\rm\small
We consider the problem of whether there are deterministic theories 
describing the evolution of an individual physical system in terms
of the definite trajectories of its constituent particles and
which stay in the same relation to Quantum Mechanics as Bohmian
Mechanics but which differ from the latter for what concerns the
trajectories followed by the particles. Obviously, one has to impose on the
hypothetical alternative theory precise physical requirements. We
analyse various such constraints and we  show step by step how to meet
them. This way of attacking the problem turns out to be useful also
from a pedagogical point of view since it allows to recall and focus
on some relevant features of Bohm's theory. One of the central requirements
we impose on the models we are going to analyse has to do with their
transformation properties under the transformations of the extended Galilei
group. In a context like the one we are interested in one can put
forward various requests that we refer to as physical and genuine
covariance and invariance. Other fundamental requests are that the theory
allows the description of isolated physical systems as well as that it
leads to a solution (in the same sense as Bohmian Mechanics) of the
measurement problem.We show that, even when all above conditions are taken
into account, there are infinitely many inequivalent (from the point of
view of the trajectories) bohmian-like theories reproducing the predictions
of Quantum Mechanics. This raises some
interesting questions about the meaning of Bohmian Mechanics.

\normalsize
\section{Introduction}
The reasons which have led David Bohm to the formulation of his
deterministic hidden variable theory have been expressed with great
lucidity by John Bell in \it Against Measurement \rm\cite{JB1}:

``In the beginning, Schr\"{o}dinger tried to interpret his
wavefunction as giving somehow the density of stuff of which the
world is made. He tried to think of an electron as represented by a
wavepacket ... a wavefunction appreciably different from zero only
over a small region in space. The extension of that region he
thought of as the actual size of the electron ... his electron was
a little bit fuzzy. At first he thought that small wavepackets,
evolving according to the Schr\"{o}dinger equation, would remain
small. But that was wrong. Wavepackets diffuse, and with the
passage of time become indefinitely extended, according to the
Schr\"{o}dinger equation. But however far the wavefunction has
extended, the reaction of a detector to an electron remains spotty.
So Schr\"{o}dinger's realistic interpretation did not survive.

Then came the Born interpretation. The wavefunction gives not the
density of stuff, but gives rather (on squaring its modulus) the
density of probability. Probability of what, exactly? Not of the
electron being there, but of the electron being found there, if its
position is ``measured".

Why this aversion to ``being" and insistence on ``finding"? The
founding fathers were unable to form a clear picture of things on
the remote atomic scale. They became very aware of the intervening
apparatus, and of the need for a ``classical" base from which to
intervene on the quantum system. And so the shifty split.

The kinematics of the world, in this orthodox picture is given by a
wavefunction (maybe more than one?) for the quantum part, and
classical variables ... variables which have values ... for the
classical part:

\begin{center}
$(\Psi(t,q,...),X(t),...)$
\end{center}

The X's are somehow macroscopic. This is not spelled out very
explicitly. The dynamics is not very precisely formulated either.
It includes a Schr\"{o}dinger equation for the quantum part, and
some sort of classical mechanics for the classical part, and
``collapse'' recipes for their interactions.

It seems to me that the only hope of precision with this dual
$(\Psi,x)$ kinematics is to omit completely the shifty split, and
let both $\Psi$ and $x(t)$ refer to the world as a whole. Then the
$x$'s must not be confined to some vague macroscopic scale, but
must extend to all scales. In the picture of de Broglie and Bohm,
every particle is attributed a position $x(t)$. Then instrument
pointers
... assemblies of particles, have positions, and experiments have
results. The dynamics is given by the world Schr\"{o}dinger
equation \it plus \rm precise ``guiding'' equations prescribing how
the $x(t)$'s move under the influence of $\Psi$''.

These sentences describe the essence of Bohm's
theory in a very lucid way. We consider it unnecessary to comment
 on its relevance and on
the crucial role it has played for the debate on the conceptual
foundations of Quantum Mechanics. As everybody knows, it allowed a
deeper understanding of some basic features of the quantum aspects
of natural phenomena. First of all, it has put into evidence the
unappropriateness of von Neumann's conclusion about the
impossibility of a deterministic completion of Quantum Mechanics.
Secondly, it has been the main stimulus for Bell's investigations
which have compelled the scientific community to face the nonlocal
aspects of nature.

It is the purpose of this paper to investigate whether the bohmian
program of reproducing the predictions of Quantum Mechanics within
a framework in which particles have definite trajectories leads
unavoidably, when some necessary general requirements are taken
into account, to Bohmian Mechanics, or if other theories exhibiting
the same features can be devised. The conclusion will be that there
exists actually infinitely many such theories. This poses some
interesting problems for the bohmian theory itself.

\section{A short survey of Bohmian Mechanics and its more relevant features}

Bohmian theory [2,3,4] is a hidden variable theory describing the
states of individual physical systems in terms of their
wavefunction $\Psi(\bf q_{\rm 1},\bf q_{\rm 2},...,\bf q_{\it
N};\it t)$ in configuration space and the actual positions $\bf
Q_{\it k}(\it t)$ of all particles of the system. The rules of the
game are quite simple:
\begin{itemize}
\item Assign $\Psi(\bf q_{\rm 1},\bf q_{\rm 2},...,\bf q_{\it N};\rm 0)$ and
$\bf Q_{\it k}(\rm 0)$.
\item Consider the two evolution equations:
\begin{eqnarray}\label{2.1}
&&\spc\spc i\hbar\frac{\partial\Psi(\bf q_{\rm 1},\bf q_{\rm
2},...,\bf q_{\it N};\it t)}{\partial t}=\hat H\Psi(\bf q_{\rm
1},\bf q_{\rm 2},...,\bf q_{\it N};\it t)\\ &&\spc\spc\left.
\frac{d\bf Q_{\it k}(\it t)}{dt}=\frac{\hbar}{m_k}
\Im\frac{\Psi^{\ast}(\bf q_{\rm 1},q_{\rm 2},...,q_{\it N};\it t)\nabla_{k}
\rm\Psi(\bf q_{\rm 1},q_{\rm 2},...,q_{\it N};\it t)} {|\rm\Psi(\bf q_{\rm
1},\bf q_{\rm
2},...,\bf q_{\it N};\it t)|^{\rm 2}}\right |_{\bf q_{\it i}=\bf
Q_{\it i}}
\end{eqnarray}

\item Solve (2.1) for the initial condition $\Psi(\bf q_{\rm 1},\bf q_{\rm
2},...,\bf q_{\it N};\rm 0)$, insert the solution in the r.h.s.\ of (2.2)
and solve
it for the initial conditions $\bf Q_{\it k}(\rm 0)$. In this way
one uniquely determines $\bf Q_{\it k}(\it t)$.
\end{itemize}

The fundamental feature of the theory consists in the fact that if
consideration is given to an ensemble of physical systems (all described by
the same wavefunction $\Psi(\bf q_{\rm 1},q_{\rm
2},...,q_{\it N};{\it t})$) each
containing \it N \rm particles whose positions are\footnote{From now on we
will similarly use the shorthand $\bf q$ for $\bf
(q_{\rm1},q_{\rm2},...,q_{\it N})$.} $\bf Q=(Q_{\rm1},Q_{\rm2},
...,Q_{\it N})$ (evolving as prescribed by Eq.\ (2.2)) and such that the
probability density $\rho(\bf Q,\it t)$ for the configuration \bf Q \rm
satisfies:
\begin{equation}\label{2.3}
\rho(\bf Q,\rm 0)=|\Psi (\bf Q,\rm 0)|^2
\end{equation}
then, the constituent particles of the systems of the ensemble
follow trajectories such that one has, at all times:
\begin{equation}\label{2.4}
\rho(\bf Q,\it t)=|\rm\Psi(\bf Q,\it t)|^{\rm 2}.
\end{equation}

The fact that Eq.\ (2.3) implies Eq.\ (2.4) is referred to as \it
equivariance \rm of the functional $\rho^{\Psi}=|\Psi|^2$. One can
easily prove it as follows. If one considers the dynamical law specified
 by the velocity field\footnote{Eq.\
(2.5) is meaningful only if the velocity field is assumed to be
one-valued and to satisfy appropriate regularity properties which
we will discuss below.} $\bf v_{\it Bk}\rm(\bf Q\rm;\it t)$\rm:
\begin{equation}\label{2.5}
\frac{\it d\bf Q_{\it k}(\it t)}{\it dt}=\bf v_{\it Bk}\rm(\bf
Q\rm;\it t);
\end{equation}
then, as is well known, the probability density $\rho(\bf Q\rm;\it t)$
for the configuration \bf Q \rm obeys the continuity equation
\begin{equation}\label{2.6}
\frac{\partial\rho(\bf Q\rm ;\it t)}{\it \partial t}+\sum_{\it
k=\rm 1}^{\it N}\nabla_{\it k}\cdot[\rho(\bf Q\rm ;\it t)\bf v_{\it
Bk}\rm(\bf Q\rm;\it t)]=\rm 0.
\end{equation}
There follows that if we choose
\begin{equation}\label{2.7}
\bf v_{\it Bk}\rm(\bf Q\rm;\it t)=\frac{\bf j_{\it QMk}\rm(\bf Q\rm;\it t)}
{|\rm\Psi(\bf Q;\it t)|^{\rm 2}},
\end{equation}
where $\bf j_{\it QMk}\rm(\bf Q\rm;\it t)$ is the quantum
mechanical current density for particle \it k\rm :
\begin{equation}\label{2.8}
\bf j_{\it QMk}\rm(\bf Q\rm;\it t)=\frac{\hbar}{m_k}\Im[\rm\Psi^\ast(\bf
Q\rm;\it t)\nabla_k\rm\Psi(\bf Q\rm;\it t)],
\end{equation}
then the solution of equation (2.6) (given the initial condition
(2.3)) is $\rho(\bf Q,\it t)=|\rm\Psi(\bf Q,\it t)|^{\rm 2}$ as it
is easily seen by taking into account the quantum continuity
equation:
\begin{equation}\label{2.9}
\frac{\partial|\Psi(\bf Q\rm ;\it t)|^{\rm 2}}{\it \partial t}+\sum_{\it
k=1}^{\it N}\nabla_{\it k}\cdot[\bf j_{\it QMk}\rm(\bf Q\rm;\it
t)]=\rm 0.
\end{equation}

Before concluding this brief account of Bohmian Mechanics we would
like to recall some relevant points:
\begin{list}{-}{}
\item In the bohmian spirit, Standard Quantum Mechanics (SQM) is
considered as an incomplete theory. To describe an individual
system, besides its wavefunction one needs further (hidden)
variables: the positions! The repetition of the preparation
procedure leading to $\Psi(\bf q;\rm 0)$ leads to a distribution of the hidden
 variables
such that (2.3) holds. The precise trajectories followed by the
particles are such to reproduce at all times the distribution of
the outcomes of position measurements predicted by SQM.
\item Each particle of the universe has at all times a precise position.
\item No other assumption is made: there are no measurements, no
wavepacket reductions and so on.
\item There are contextual aspects giving rise to problems of
remarkable epistemological relevance if one pretends that knowledge of the
hidden variables determines the precise value of any conceivable observable
of the standard formulation. All such problems are overcome simply by
taking the attitude that what the theory is about are the positions and
the trajectories of all particles and nothing else. With reference
to the remarks by Bell in the above quotation, the theory claims
that $|\Psi(\bf Q\rm;\it t)|^{\rm 2}$ gives the probability density
of the configuration \it being \bf Q \rm at time \it t \rm and not, as SQM, the
probability density of \it finding \rm the outcome \bf Q \it if a
measurement is performed\rm.
\end{list}

We would also like to point out that, recently, there have been
extremely interesting investigations \cite{DGZ} aimed to clarify
the so called Quantum Equilibrium Hypothesis, i.e.\ in which sense
the initial condition $\rho(\bf Q,\rm 0)=|\Psi (\bf Q,\rm 0)|^2$
comes about. Note that if this condition is not satisfied by the
members of the ensemble and bohmian evolution equations hold, then
the actual distribution at time \it t \rm would contradict the
predictions of SQM. Similarly, there have been detailed analysis
\cite{BDG} proving the uniqueness and the global existence of the
solutions of the dynamical equation (2.5) for the trajectories, a
highly nontrivial problem due, e.g., to the possible vanishing of
the wavefunction or to its behaviour at infinity. It is important
to stress that such investigations refer to the theory as it has
been presented, i.e.\ with the prescription that the velocity field
is precisely the one defined above in terms of the wavefunction.

This last remark has to be kept in mind to fully appreciate the
problem to which the present paper is addressed: we will not
question the uniqueness of the solutions for a given velocity
field, but of the velocity field itself, i.e.\ we will face the
problem of \it whether there are theories which are equivalent to
SQM in the same sense of Bohmian Mechanics, but which describe the
evolution in terms of different trajectories\rm.

\section{Is Bohmian Mechanics Unique?}
As just mentioned, the question we are interested in is the
following. Suppose we devise a theory which completes Quantum
Mechanics by exhibiting the feature that all particles follow
definite trajectories and thus have at any time definite positions.
The only allowed ingredients of the theory are the wavefunction and
the particle positions. Does such a request lead unavoidably to
Bohmian Mechanics? That this is not the case can be shown in a very
elementary way. One keeps the statevector and the Schr\"{o}dinger
evolution equation for it and one changes the dynamical equation
for the trajectories by adding to the bohmian velocity term $\bf
v_{\it Bk}\rm(\bf Q\rm;\it t)$ an additional velocity field $\bf
v_{\it Ak}\rm(\bf Q\rm;\it t)$ given by:
\begin{equation}\label{3.1}
\bf v_{\it Ak}\rm(\bf Q\rm;\it t)=\frac{\bf j_{\it Ak}\rm(\bf Q\rm;\it t)}
{|\rm\Psi(\bf Q;\it t\rm)|^2},
\end{equation}
where $\bf j_{\it Ak}\rm(\bf Q\rm;\it t)$ satisfies:
\begin{equation}\label{3.2}
\sum_{k=1}^{N}\nabla_k\cdot\bf j_{\it Ak}\rm(\bf Q\rm;\it t\rm)=0
\end{equation}
One can then argue as follows. The equation:
\begin{equation}\label{3.3}
\frac{\it d\bf Q_{\it k}(\it t)}{\it dt}=\bf v_{\it Bk}\rm(\bf
Q\rm;\it t)+\bf v_{\it Ak}\rm(\bf Q\rm;\it t)
\end{equation}
implies
\begin{equation}\label{3.4}
\frac{\partial\rho(\bf Q\rm ;\it t)}{\it \partial t}+\sum_{\it
k=1}^{\it N}\nabla_{\it k}\cdot\{\rho(\bf Q\rm ;\it t)[\bf v_{\it
Bk}\rm(\bf Q\rm;\it t)+\bf v_{\it Ak}\rm(\bf Q\rm;\it t)]\}=\rm 0.
\end{equation}
It is then obvious that the function
\begin{equation}\label{3.5}
\rho(\bf Q,\it t)=|\rm\Psi(\bf Q;\it t)|^{\rm 2}
\end{equation}
is the solution of Eq.\ (3.4) satisfying the condition $\rho(\bf
Q,\rm 0)=|\Psi(\bf Q,\rm 0)|^2$ as a consequence of the SQM
continuity equation since, with the choice (3.5), the last term of
Eq.\ (3.4) vanishes in virtue of (3.2). We stress that Eq.\ (3.3)
implies that, in general, the trajectories are different from those
of Bohmian Mechanics which are characterised by $\bf v_{\it
Ak}\rm(\bf Q\rm;\it t)=\rm 0$.

Obviously, one must be particularly careful in choosing the new
velocity field to guarantee the existence, uniqueness and globality
of the solution. Even more, the regularity conditions for the velocity field
which make legitimate to pass from Eq.\ (3.3) to (3.4) could fail
to hold. These problems will be dealt with in the next Sections.

We mention that E.~Squires \cite{SQ} has considered a modification
of this type in which $\bf j_{\it Ak}\rm(\bf Q\rm;\it t)$ was
simply a constant current. Unfortunately, such a choice implies that there
are particles which flow in and out at infinity, so
that particles can be ``created'' and/or ``destroyed'', a
physically unacceptable fact in the present context. Moreover, the
theory turns out to be non-covariant (see below). For these reasons
Squires himself has considered unviable the line of thought we are
pursuing here.

\section{Sufficient conditions for the velocity field}
The considerations of the final part of the previous Section could
lead one to think that the necessary mathematical and physical
requirements which have to be imposed to the theory, would reduce
the family of possible dynamical models we are envisaging to
Bohmian Mechanics itself. As we shall show, this is not the case,
but, on the contrary, there are infinitely many sensible bohmian-like model
theories with different trajectories.

For what concerns the mathematical properties of the velocity field
which guarantee the existence of the trajectories and the validity
of the continuity equation, we do not want to be too technical. We
will limit ourselves to point out that if the additional velocity field is
defined everywhere and continuous
and not diverging more than linearly at infinity, no
particle can reach infinity in a finite time, the flux vanishes at infinity
and thus the model is physically sensible. In refs.\ \cite{BDG} the
implications of the occurrence of possible singularities of the velocity
field in the standard bohmian framework have been discussed and proved to
have no unacceptable consequences. Here we will not go through such a
detailed mathematical analysis of the properties of the additional velocity
fields and we will suppose that the expressions we will introduce do not
give rise to any problem.

In the next Section we start by imposing necessary physical requests on
the models we are envisaging by analysing the delicate
problem, within a context like ours, of the appropriate covariance
and invariance requirements for the theory.

\section{Covariance and Invariance: Genuine versus Physical}
It is important to focus on the particular aspects
that the problem of covariance acquires in a context like Bohmian
Mechanics. We will obviously deal with the transformations of the
Galilei group. There are different requests that one can put
forward and that we will denote as genuine and physical covariance
requirements. In fact, one should keep clearly in mind, as shown in ref.\
[5], that as a consequence of absolute uncertainty, the hidden variables of
the theory are unaccessible. Thus, if one considers as
 meaningful only
what the theory claims to be physically accessible, one could be
satisfied with the request that different observers agree on the
``objective'' probability density for the various configurations at
a given time. Since the class of theories we are envisaging are
assumed to give configuration probability densities in agreement
with those of Standard Quantum Mechanics, such ``physical'' covariance
conditions turn out to be automatically satisfied.

However, as Bell \cite{JB2} has appropriately stressed, for a theory like
the one under consideration one should impose requests which
go beyond the one of simple agreement with experiment. He demands
that the theory does not admit any (even hidden) preferred
reference frame. In simpler words, what Bell requires is that
observers related to one another by a transformation of the Galilei
group agree on the (unaccessible) trajectories that the particles
follow. If this happens the theory is said to be genuinely
covariant.

In the case in which the underlying quantum problem exhibits
invariance under the transformations of the Galilei group, a
similar distinction could be made with reference to the
bohmian-like models we are considering. In particular, in such a
case, the request of genuine invariance amounts to demanding that
observers connected by a transformation of the Galilei group
which consider systems in the ``same'' --- in their own language ---
initial conditions (thus corresponding to objectively different
physical situations) will agree on the trajectory that each
individual particle follows.

In our case, since the Schr\"{o}dinger equation already satisfies
the covariance and/or invariance requests, we can limit our
considerations to the properties of the velocity fields determining
the trajectories. We will discuss in detail only the case of
special Galilei transformations and we will mention the
transformation properties of the velocity field for the other
transformations of the Galilei group. As we will see, Bohmian
Mechanics in its standard form is physically and genuinely
covariant and, consequently, invariant if the corresponding quantum
problem is invariant under the transformations of the Galilei
group.

\section{Special Galilei Transformations}
Let us discuss, first of all, how the bohmian velocity field $\bf
v_{\it Bk}(Q;\it t)$ changes under a special Galilei transformation:
\begin{equation}\label{6.1}
\left\{\begin{array}{l}
\bf q'_{\it i}=q_{\it i}-v\it t \\
\it t' =t.
\end{array}\right.
\end{equation}
To this purpose let us suppose that an observer O considers a
specific initial wavefunction $\Psi(\bf q,\rm 0)$ and a specific
configuration $\bf Q(\rm 0)$. By using Eq.\ (2.1) and the precise
prescription (2.7) of the theory, he determines the wavefunction
$\Psi(\bf q,\it t)$ and the velocity field $\bf v_{\it Bk}(Q;\it
t)$. In turn, such a field, when inserted into (2.5), uniquely
determines the trajectory $\bf Q(\it t)$. Another observer $\rm O'$
considers the same \it objective \rm wavefunction\footnote{Obviously, in
the case of an arbitrary transformation of the group in place of Eq.\
(6.1), the functional dependence of $\Psi'$ from its argument will be
different from the one of $\Psi$. Note, in fact, that the specific relation
between $\Psi$ and $\Psi'$ derives also from the fact that, according to
Eq.\
(6.1), for $t=0$ $\bf q'=q$.} in his language
$\Psi'(\bf q';\rm 0)=\Psi(\bf q';\rm 0)$ and uses
the same formal procedure as the original observer (i.e.\ it leaves
$\Psi'(\bf q';\rm 0)$ evolve according to \it his \rm Hamiltonian,
which may be different from the one of the original observer),
evaluates the corresponding velocity field $\bf v'_{\it Bk}(Q';\it
t')$ and studies the evolution of the particle which initially is
at the same \it objective \rm point.

We recall \cite{FG} that the Galilei transformation (6.1) is
implemented by the unitary operator
\begin{equation}\label{6.2}
{\cal G}_{\bf v}(\it t)=\rm e^{\it-\frac{i}{\rm 2\it\hbar}Mv^{\rm 2}\it t}
e^{-\it\frac{i}{\hbar}M\hat{\bf Q}\cdot \bf v}
e^{\it\frac{i}{\hbar}\hat{\bf P}\cdot\bf v\it t},
\end{equation}
$M=\sum_{i=1}^{N}m_i$ being the total mass and $\bf
\hat Q,\hat P$ the centre-of-mass position and momentum
operators, respectively.

Equation (6.2) shows that the wavefunctions for $\rm O'$, in his
variables $\bf q'_{\it i}$ and $t'$, is:
\begin{equation}\label{6.3}
\Psi'(\bf q';\it t')=\langle\bf q'|{\cal G}_{\bf v}(\it t')|\rm\Psi,\it
t\rangle=
\rm e^{\it-\frac{i}{\rm 2\it\hbar}Mv^{\rm 2}\it t} e^{-\it\frac{i}{\hbar}M\bf q
'\cdot v}
\Psi(\bf q'_{\it i}+v\it t',t').
\end{equation}
As we have seen in Section 2, the observer O determines the
particle trajectories according to:
\begin{equation}\label{sei}
\frac{d\bf Q_{\it k}(\it t)}{dt}=\bf v_{\it Bk}(\bf Q;\it t),\;\;\;
\bf v_{\it Bk}(\bf Q;\it t)=\left. \frac{\hbar}{m_k}
\Im\frac{\rm\Psi^{\ast}(\bf q;\it t)\nabla_{k}
\rm\Psi(\bf q;\it t)} {|\rm\Psi(\bf q;\it t)|^{\rm 2}}\right |_{\bf q=Q}
\end{equation}
and the observer $\rm O'$ determines his own velocity field $\bf
v'_{\it Bk}(\bf Q';\it t')$ by using the corresponding
prescription, getting:
\begin{eqnarray}
&&\spc\bf v'_{\it Bk}(Q';\it t')=\left. \frac{\hbar}{m_k}
\Im\frac{\rm\Psi'^{\ast}(\bf q';\it t')\nabla'_{k}
\rm\Psi'(\bf q';\it t')} {|\rm\Psi'(\bf q';\it t')|^{\rm 2}}\right |_{\bf
q'=Q'}\nonumber\\
&&\spc =\left.\frac{\hbar}{m_k}\Im\frac{\Psi(\bf q'_{\it i}+v\it
t',t')\;\nabla'_k\rm\Psi(\bf q'_{\it i}+v\it t',t')}{|\rm\Psi(\bf
q';\it t')|^{\rm2}}\right |_{\bf
q'=Q'}+\it\frac{\hbar}{m_k}\Im\left(-\frac{i}{\hbar}m_k\bf v\right
)\nonumber\\ &&\spc =\bf v_{\it Bk}(\bf Q'+v'\it t';t')-\bf v.
\end{eqnarray}
Since $\rm O'$ uses the equation:
\begin{equation}\label{6.6}
\frac{d\bf Q'_{\it k}(\it t')}{dt'}=\bf v'_{\it Bk}(\bf Q';\it t')=
\bf v_{\it Bk}(\bf Q'+v\it t';t')-\bf v,
\end{equation}
if $\bf Q_{\it k}(\it t)$ satisfies the first of Eqs.\ (\ref{sei}),
then
\begin{equation}\label{6.7}
\bf Q'_{\it k}(\it t')=\bf Q_{\it k}(\it t)-\bf v\it t
\end{equation}
satisfies (6.6).

This proves the genuine covariance of the trajectories. Obviously,
when the underlying quantum theory is invariant for the
transformation considered, the Hamiltonian for $\rm O'$ coincides with the
one for O and the above procedure shows that if one considers the
same\footnote{Here, by the expression ``the same'', we mean that
the functional dependence of the initial wavefunction and the
numerical value of the initial position is the same for the two
observers. Note that this means that they are considering two
``objectively different'' physical situations.} wavefunction and
the same initial conditions in the two reference frames, one gets
exactly the same trajectories.

The conclusion is obvious: Bohmian Mechanics exhibits genuine
covariance and/or invariance for special Galilei transformations.
We simply list here the transformation properties which
characterise the bohmian velocity field for all the
transformations of the extended Galilei group:
\begin{equation}
\begin{array}{ll}
\mbox{i. Space translations:} & \bf v'_{\it Bk}(\bf Q';\it t')=\bf v_{\it
Bk}(\bf Q'+A;\it t')\\
\mbox{ii. Time translations:} & \bf v'_{\it Bk}(\bf Q';\it t')=\bf v_{\it
Bk}(\bf Q';\it t'+\tau)\\
\mbox{iii. Space rotations:} & \bf v'_{\it Bk}(\bf Q';\it t')={\cal R}\bf
v_{\it Bk}({\cal R}^{\rm-1}\bf Q';\it t')\\
\mbox{iv. Galilei boosts:} & \bf v_{\it Bk}(\bf Q';\it t')=\bf v_{\it
Bk}(\bf Q'+v\it t'; t')-\bf v \\
\mbox{v. Space reflection:} & \bf v'_{\it Bk}(\bf Q';\it t')=-\bf v_{\it
Bk}(-\bf Q';\it t')\\
\mbox{vi. Time reversal:} & \bf v'_{\it Bk}(\bf Q';\it t')=-\bf v_{\it
Bk}(\bf Q';-\it t').
\end{array}
\end{equation}
These, in turn, guarantee the genuine covariance and/or invariance
of the theory for the extended Galilei group.

\section{Looking for alternative models}
As already stated, it is possible to find a whole family of model
theories which attribute definite trajectories to the particles and,
 with respect to SQM, stay in the same relation as Bohm's theory. With
reference to the case of a system of $N$
particles and to the formalism of Section 3,
let us suppose that we choose for the additional velocity field
$\bf v_{\it Ak}(Q;\it t)$ the following expression:
\begin{eqnarray}\label{7.1}
\bf v_{\it Ak}(\bf Q;\it t)&=&\gamma_{\it k}
\left.\frac{\nabla_k\times
\{|\rm\Psi(\bf q;\it t)|^{\rm2}[\bf v_{\it Bk}
(\bf q;\it t)]\}}{\rm|\Psi(\bf q;\it t)|^{\rm2}}\right |_{\bf q=Q}\nonumber\\
&=&\left.\frac{i\hbar\gamma_k}{m_k}\frac{\nabla_{\it
k}\rm\Psi}{\rm\Psi}\times\frac{
\nabla_{\it k}\rm\Psi^*}{\rm\Psi^*}\right |_{\bf q=Q}.
\end{eqnarray}
In Eq.\ (\ref{7.1}) $\gamma_{\it k}$ is an appropriate dimensional constant
(we can choose it to be $[\hbar/m_{\it k}c]$ without introducing any
new constant of nature in the theory) and $\nabla_k
\times$ denotes the curl differential operator over the coordinates of the
\it k\rm -th particle.
The velocity field (\ref{7.1}) is easily proved to exhibit the same
transformation properties (6.8i)--(6.8iii) of the bohmian
velocity field. On the contrary, its transformation properties under
Galilei boosts are such to violate the covariance requirements for
the theory. To take seriously the model we are proposing we have
then, first of all, to overcome such a difficulty. As a first attempt to
reach this goal, let us start by considering a system of $N$ spinless
particles interacting via a potential which depends only on the
modulus of the difference of their coordinates so that one can be tempted
to adopt the following line of thought.
As is well known, in such a case the bohmian dynamics is equivalent to the one
obtained by resorting to the centre-of-mass (\bf R\rm) and relative
($\bf r_{\it k}$\rm) coordinates:
\begin{eqnarray}\label{7.5-8}
&&\spc i\hbar\frac{\partial\Psi(\bf R,r_{\it k};\it t)}{\partial t}=\hat
H\Psi(\bf R,r_{\it k};\it t),\label{7.5}\\ &&\spc\hat H=\frac{\bf\hat P\rm
^2}{2M}+\sum_{\it k}\frac{\bf\hat p_{\it k}\rm^2}{2\mu_{\it k}}+V,\label{7.6}\\
&&\spc\left.
\frac{d\bf\tilde R(\it t)}{dt}=\frac{\hbar}{M}
\Im\frac{\Psi^{\ast}(\bf R,r_{\it k};\it t)\nabla_{R}
\rm\Psi(\bf R,r_{\it k};\it t)} {|\rm\Psi(\bf R,r_{\it k};\it t)|^{\rm 2}}
\right |_{\bf R=\tilde R,r_{\it k}=\tilde r_{\it k}}=\bf v_{\it BR}(\tilde
R,\tilde r_{\it k};\it t),\label{7.7}\\
&&\spc\left.
\frac{d\bf\tilde r_{\it k}(\it t)}{dt}=\frac{\hbar}{\mu_{\it k}}
\Im\frac{\Psi^{\ast}(\bf R,r_{\it k};\it t)\nabla_{rk}
\rm\Psi(\bf R,r_{\it k};\it t)} {|\rm\Psi(\bf R,r_{\it k};\it t)|^{\rm 2}}
\right |_{\bf R=\tilde R,r_{\it k}=\tilde r_{\it k}}=\bf v_{\it Bk}(\tilde
R,\tilde r_{\it k};\it t),\label{7.8}
\end{eqnarray}
where we have
denoted by $M$ and $\mu_{\it k}$ the total and reduced masses respectively.

This fact suggests to overcome the difficulty by keeping the first three of
the above equations and changing the remaining ones by adding to the
bohmian velocity fields of the relative motions new terms analogous to the
one of Eq.\ (\ref{7.1}), so that Eqs.\ (7.5) are replaced by:
\begin{equation}\label{7.9}
\frac{d\bf\tilde r_{\it k}(\it t)}{dt}=\bf v_{\it Bk}(\tilde R,\tilde
r_{\it k};\it t)+
\bf v_{\it Ak}(\tilde R,\tilde r_{\it k};\it t)\equiv \bf v_{\it Nk}(\tilde
R,\tilde r_{\it k};\it t),
\end{equation}
where
\begin{equation}\label{7.10}
\bf v_{\it Ak}(\tilde R,\tilde r_{\it k};\it t):=\left.\gamma_{\it k}
\frac{\nabla_{rk}\times
[|\rm\Psi(\bf R,r_{\it k};\it t)|^{\rm 2} \bf v_{\it Bk}(R,r_{\it k};\it
t)]}{\rm|\Psi(\bf R,r_{\it k};\it t)|^{\rm 2}}\right |_{\bf R=\tilde
R,r_{\it k}=\tilde r_{\it k}}.
\end{equation}
The velocity fields of the just considered model obviously have the correct
behaviour under the transformations of the proper Galilei group and, in
virtue of Eq.\ (\ref{7.9}) the trajectories of the individual particles
differ from those of Bohmian Mechanics.

Up to now we have not considered the transformations of the extended group.
It is easily seen that the velocity fields of Eq.\ (\ref{7.10}) transform
like pseudovectors under space reflection so that the theory is not
covariant for the full group. One could then introduce quantities like
\begin{equation}\label{7.11}
\spc f_{\it k}^{\Psi}(t):=i\frac{d}{dt}\int
d^{3(\it N\rm -1)}rd^3R\left\{[\nabla_k\Psi(\bf R,r_{\it k};\it
t)\times\nabla_{\it
k}\rm\Psi^{*}(\bf R,r_{\it k};\it t)]\cdot\frac{\bf r_{\it k}}{\it r_{\it
k}}\right\}.
\end{equation}
Such functions are continuous, real scalar functions of $t$, which
change sign for space reflections and do not change sign for
time reversal. Therefore, if we define new velocity fields:
\begin{equation}\label{7.12}
\bf\tilde v_{\it Ak}(\bf R,r_{\it k};\it t)=f_k^{\rm\Psi}(t)\bf v_{\it
Ak}(R,r_{\it k};\it t)
\end{equation}
and we replace in all previous formulae $\bf v_{\it Ak}(R,r_{\it k};\it t)$
with $\bf\tilde v_{\it Ak}(R,r_{\it k};\it t)$, we have the appropriate
transformation properties of  the additional velocities under the
transformations of the full Galilei group.

\section{Some serious drawbacks of the model of Section 7}
The model theory we have just presented seems, at first sight, physically
sensible and satisfies all covariance requirements one can put forward for it.
Obviously, the model exhibits a formal feature which could be considered as
not fully satisfactory, i.e. its physical implications (more precisely the
precise trajectories followed by the various particles) depend on the
particular choice one makes for the relative coordinates. Said differently,
if one denotes by $A$ a precise transformation leading from the absolute
$\{\bf q_{\it i}\}$ to the relative $\{\bf R, r_{\it i}\}$ coordinates, and
one resorts to Eqs. (7.2--7.4) and (7.6--7.7) to define the dynamics, the
evolution of another set of relative variables $\{\bf R, \hat r_{\it i}\}$
related
to $\{\bf q_{\it i}\}$ by a different transformation $\hat A$ is governed
by velocity fields which have not the formal structure (\ref{7.10}). This,
by itself, does not represent a drawback of the model\footnote{Actually,
once the precise choice for the relative variables is made, the theory is
consistently and uniquely defined and turns out to be Galilei covariant and
could be expressed in terms of the absolute coordinates and the associated
derivatives, avoiding even to mention the relative
variables.}; rather one could see it as a further direct and simple proof
that one can exhibit many different theories of the kind we are looking
for.

However, if one analyses in a more critical way the proposed model, one
easily realises that its characteristic dependence on the choice made for
the relative variables, beside making it formally ``less elegant'' than
Bohmian Mechanics, has some unacceptable physical implications. Suppose in
fact that the particles of the physical system under consideration can be
arranged in various groups, such that the members of different groups are
non-interacting and non-entangled (this last specification referring to the
statevector of the system in the absolute coordinates). In such a case SQM as
well as Bohmian Mechanics allow one to consider the physical systems
composed by the particles of any group as isolated from the others (at
least until the evolution brings particles of different groups into
interaction). Accordingly, the evolution of the particles of one group does
not depend in any way whatsoever on the members of other groups. If this
does not happen, one could legitimately claim that the model exhibits
``more'' nonlocal aspects than those which are unavoidably brought into
play by quantum entanglement.

It should be evident that the model of the previous Section does actually
violate the requirement we have just put forward. In fact, if one writes it
in the absolute coordinates, one sees that the additional velocity
corresponding to the variable $\bf q_{\it k}$ is, in general, a linear
combination of terms of the type\footnote{Again, as stated in footnote 1,
we use \bf q \rm as a shorthand for $(\bf q_{\rm 1},q_{\rm 2},...,q_{\it
N})$.} $\displaystyle{\frac{\nabla_{\it
l}\Psi(\bf q;\it t)}{\rm\Psi(\bf q;\it t)}\times\frac{\nabla_{\it j}\Psi^*(\bf
q;\it t)}{\rm\Psi^*(\bf q;\it t)}}$ involving also particles $l$ and $j$
(in which $l$ or $j$ might coincide with $k$) which could be
non-interacting and non-entangled with the $k$-th particle.

Concluding, the fundamental request that one can deal with isolated systems
(a possibility which, in Bell's words \cite{JB1}, has marked the birth of
experimental science) is not met by the model based on Eq.\ (\ref{7.10}),
so that it turns out to be physically unacceptable. Let us analyse whether
our program can still be pursued.

One can devise various ways to circumvent the just identified difficulties.
Let us begin by considering some elementary cases limiting (for the moment)
our considerations to models which are covariant only under the
transformations of the proper group. There are two possible strategies one
can follow. The first consists in building expressions for the additional
velocity field associated to a given absolute coordinate which involve only
derivatives with respect to the same coordinate (contrary to what happens
for the previous model) and which, moreover, in the case in which the
wavefunction in the absolute coordinates factorizes in two (or more)
factors, depend only on the factor to which the coordinate under
consideration belongs (this typically happens for expressions like
$\displaystyle{\frac{\nabla_{\it k}\Psi(\bf q;\it t)}{\rm\Psi(\bf q;\it
t)}}$, $k$
being the considered particle).
Another strategy allows the appearance of derivatives referring also to
other variables, provided they are entangled with the one under
consideration.

To be more specific, let us list some cases. From now on, we will always
make use of the language of the absolute coordinates. The general
philosophy, as we know, is that of adding to the r.h.s.\ of Eq.\ (2.2)
an additional velocity term as shown in Eq.\ (\ref{3.3}). Let us consider
various alternatives:
\begin{eqnarray}\label{8.1-3}
\bf v_{\it Ak}^{\rm (1)} (\bf q ;\it t)\propto\rm\frac{\nabla_{\it
k}|\Psi(\bf
q ;\it t)|^{\rm 2}}{\rm|\Psi(\bf q ;\it t)|^{\rm 2}}\times\frac{\it d\bf
v_{\it Bk}(\bf q
;\it t)}{\it dt}\\
\bf v_{\it Ak}^{\rm (2)} (\bf q ;\it t)\propto\rm\frac{\nabla_{\it
k}|\Psi(\bf
q ;\it t)|^{\rm 2}}{\rm|\Psi(\bf q ;\it t)|^{\rm 2}}\times\frac{\it d^{\rm
2}\bf v_{\it
Bk}(\bf q ;\it t)}{\it dt^{\rm 2}}\\
\bf v_{\it Ak}^{\rm (3)} (\bf q ;\it t)\propto\rm\frac{\nabla_{\it
k}|\Psi(\bf
q ;\it t)|^{\rm 2}}{\rm|\Psi(\bf q ;\it t)|^{\rm 2}}\times\nabla_{\it
k}^{\rm 2}\bf
v_{\it Bk}(\bf
q ;\it t)
\end{eqnarray}
where $\displaystyle{\frac {d}{dt}}$ is the following operator:
\begin{equation}\label{8.4}
\frac {d}{dt} := \frac {\partial}{\partial t}+\sum_{\it k}\bf v_{\it
Bk}\cdot\nabla_{\it k}.
\end{equation}
With a little algebra it is easy to check that $\nabla_{\it k}\cdot
[|\Psi(\bf q)|^ {\rm 2}\bf v_{\it Ak}^{\it (r)}]=\rm 0$ $(r=1,2,3)$, so that
Eq.\ (3.2) holds. The introduction of the time derivatives in Eqs.\ (8.1)
and (8.2) as well
as the one of the Laplace operator in Eq.\ (8.3) serves the purpose of
guaranteeing that $\bf v_{\it Ak}$ transforms correctly under Galilei
boosts.

To exhibit a typical example of the second approach mentioned above, we
begin by remarking that one can easily build real, positive, bounded by 1
and continuous functions $w_{\it kj}$ with
the property of vanishing whenever particle $j$ is not entangled with
particle $k$ and which involve only the coordinate of those particles which
are entangled with the $k$-th particle. Typically an expression of the type
\begin{equation}\label{8.5}
w_{\it kj}(\bf q ;\it t)=\rm
1-\exp\left\{-\gamma^{\rm2}\left[\nabla_k\cdot\nabla_j \log|\rm\Psi(\bf
q;\it
t)|\right]^{\rm2}\right\}
\end{equation}
(where $\gamma:=\frac{\hbar}{c\sum_{\it k}m_{\it k}}$) has the desired
features. One could then choose, e.g., for the additional
velocity field of the $k$-th particle the following expression:
\begin{equation}\label{8.6}
\bf v_{\it Ak}^{\rm (4)} (\bf q;\it t)\propto\frac{\nabla_{\it
k}\times\left\{\rm|\Psi|^{\rm 2}\left [\bf v_{\it Bk}-\sum_{\it j}\left
(\frac{\it w_{\it kj}}{\sum_{\it r}\it w_{\it kr}}\right ) \bf v_{\it
Bj}\right ]\right\}} {\rm|\Psi|^{\rm 2}}.
\end{equation}
As it should be obvious, the coefficients in the round brackets serve the
purpose of making $\bf v_{\it Ak}^{(\rm 4)}$ independent of the coordinates
of the particles which are not entangled with the $k$-th one and the
difference of bohmian velocities guarantees the correct transformation
properties under Galilei boosts.

The just considered additional velocities have, however, wrong
transformation properties for the improper transformations: $\bf v_{\it
Ak}^{(\rm 1)}$ changes sign for space reflections and time reversal, the
remaining ones do not transform correctly only for space reflection. One
could circumvent this drawback by a technique analogous to the one used in
Section 7, that is by multiplying $\gamma_{k}$ times a pseudoscalar factor
\begin{equation}
\spc\!\! g_{k}^{\Psi}(t)\propto\int d\bf q|\rm\Psi(\bf q;\it
t)|^{\rm2}\left\{\left[\rm\frac{\nabla_{\it k}|\Psi(\bf q;\it t)|}{\rm|\Psi(\bf
q;\it t)|}\times\nabla_{\it k}^{\rm 2}\bf v_{\it Bk}(\bf q;\it
t)\right]\cdot\frac{\it d^{\rm 2}\bf v_{\it Bk}(\bf q;\it t)}{\it dt^{\rm
2}}\right\}.
\end{equation}
At this point it seems that we have reached our goal. However
the presence of factors of the kind of $g_{k}^{\Psi}(t)$ has some
peculiar consequences we are going to discuss.

\section{The implications of integrated expressions}
The appearance of the factor $g_{k}^{\Psi}(t)$ in the expression for the
additional velocity has some peculiar consequences arising from its
involving an integral of
the wavefunction over the whole space. Such consequences deserve a detailed
analysis. We will, for simplicity's sake, confine our
considerations to the case of a single particle (even though we will
obviously have in mind a macroscopic object ---typically a pointer--- so
that the variable {\bf Q} in the following formulae can be identified with
the centre-of-mass coordinate of such an object).

Now we can discuss the peculiar aspects of the model of the previous
Section. To this purpose let us first consider the situation with reference
to the standard version of Bohmian Mechanics. There are two aspects of such
a theory which deserve a detailed analysis. The first has to do with the
role of the ``empty waves'', the second with the problem of the
identification of the {\it effective wave function} of a physical system.
Concerning the first point, suppose that the evolved of the initial
wavefunction is the sum
of two terms $\Psi(\bf q;\it t)=\rm\Psi_{\it L}(\bf q;\it t)+\rm\Psi_{\it
R}(\bf q;\it t)$ having, for the time interval $[t_0,T]$, disjoint supports
$(\Psi_{L}(\bf q;\it t)\cdot\rm\Psi_{\it R}(\bf q;\it t)=\rm 0)$. Suppose
moreover that the individual physical system we are interested in {\it is},
at time $t_0$, at a position belonging to the support of $\Psi_{L}(\bf
q;\it t)$. The guidance condition for such system in the considered time
interval will be determined, according to the theory, by the exact
wavefunction $\Psi(\bf q;\it t)=\rm\Psi_{\it L}(\bf q;\it t)+\rm\Psi_{\it
R}(\bf q;\it t)$ and by the exact differential equation
\begin{equation}\label{9.2}
\frac{\it d{\bf Q}(t)}{dt}={\bf v}_B({\bf
Q};t)=\left.\frac{\hbar}{m}\Im\frac{\nabla\Psi({\bf q};t)}{\Psi({\bf
q};t)}\right|_{\bf q=Q},
\end{equation}
${\bf v}_B({\bf Q};t)$ being the Bohmian velocity related to $\Psi({\bf
Q};t)$. Now, for the specific formal structure of Bohmian Mechanics it
happens that in the case under consideration we can replace $\Psi({\bf Q};t)$ by
$\Psi_L({\bf Q};t)$ everywhere in the dynamical equations (see for example
\cite{H}). Obviously, no wavefunction can have compact support for a finite
time interval, but the above conclusion also holds, with an accuracy which
is higher the smaller is the overlapping of the two states, when the two
terms $\Psi_L({\bf Q};t)$ and $\Psi_R({\bf Q};t)$ of the superposition have
almost disjoint supports for the time interval $[t_0,T]$ we are interested
in. In the specific case of two pointer states corresponding to two
macroscopically
different locations this holds for extremely long times.

We stress that the argument we have just developed has the following
conceptual structure: the correct dynamics of the pointer is governed by
$\Psi({\bf Q};t)$ and by Eq.\ (\ref{9.2}); however, the fact that to an
extremely high degree of accuracy one can consider the evolution as governed by
$\Psi_L({\bf Q};t)$ and by the corresponding Bohmian velocity field
$\displaystyle{{\bf v}_{BL}({\bf
Q};t)=\frac{\hbar}{m}\Im\frac{\nabla\Psi_L({\bf Q};t)}{\Psi_L({\bf Q};t)}}$
makes immediately evident that the theory solves the macro-objectification
problem, in particular that macroscopic objects like pointers in a
superposition of spatially separated states evolve as if the ``empty wave''
referring to the region not containing the pointer would not
exist\footnote{Anyway
it has to be kept in mind that the exact dynamics is determined by
$\Psi({\bf Q};t)$ and by Eq.\ (\ref{9.2}); as it is obvious, if one
considers times $t>>T$ so large that the spreading of the wave packets
leads to an appreciable overlapping, the replacement of $\Psi({\bf Q};t)$
by $\Psi_L({\bf Q};t)$ would be no more legitimate.}.

Let us analyse now the second relevant feature of Bohmian Mechanics, i.e.\
the way in which it deals with the problem of the identification of the
{\it
effective wavefunction} \cite{DGZ} of our system. In synthesis the question
can be summarized as follows: the system under consideration (as well as
an assembly of such systems) is {\it de facto} a subsystem of the (unique)
universe in which we live. In general the wavefunction of the universe at
time $t_0$ will have a form implying an entanglement of the
system we are considering and other parts of the universe itself:
\begin{equation}\label{9.1}
\tilde\Psi({\bf Q},{\bf q}_j;t_0)=\sum_s\Psi_s({\bf Q};t_0)\Phi_s({\bf
q}_j;t_0),
\end{equation}
where the wavefunctions $\Psi_s({\bf Q};t_0)$ correspond to the
macroscopically different situations of the pointer and the coordinates
${\bf q}_j$ refer to the particles of the environment of the system under
consideration.

An essential step is that of proving the legitimacy of
associating to an individual system --- the pointer with coordinate {\bf Q} (or
an ensemble of such systems) --- a specific one of
the wavefunctions $\Psi_s({\bf Q};t_0)$ of Eq.\ (9.1). The precise sense in
which this is appropriate has been discussed in great detail in a series of
beautiful papers \cite{DGZ} by D\"{u}rr et al.
The argument requires two steps: first of all, following a line of thought
analogous to the one described above in connection with the problem of the
empty wave, one proves that it is legitimate to keep only the term of Eq.\
(\ref{9.1}) corresponding to the actual location of the pointer. Secondly,
one recalls that in the case of factorized statevectors the guidance
equations for the coordinates of one of the two factors do not involve the
other factor. Thus, one can actually consider a specific $\Psi_s({\bf
Q};t_0)$ as the effective initial wavefunction of the system we are
interested in.

We can now compare the just discussed feature of the standard theory with
those of the alternative versions of it which are the subject of the
present paper. When expressions like $g_k^{\Psi}(t)$ appear, the situation
turns out to be quite different. In fact, with reference to the first of
the points discussed above, the exact guidance equation is:
\begin{equation}\label{9.3}
\frac{\it d{\bf Q}(t)}{dt}={\bf v}_B({\bf Q};t)+{\bf v}_A({\bf Q};t)
\end{equation}
showing that the intensity of the
velocity field for a particle in the support of $\Psi_L({\bf Q};t)$ depends
crucially also on the empty wave $\Psi_R({\bf Q};t)$.

The conclusion is that the appearance of integrated expressions destroys
two of the most important features of Bohmian Mechanics, i.e.\ the
possibility of disregarding the empty waves and the one of resorting to the
effective wavefunction\footnote{With reference to this second feature we
note that the appearance of integrated expressions makes illegitimate the
replacement of the superposition (9.2) with one of its terms, it does not
alter the fact that in the case of a factorized wavefunction one can
disregard one of the two factors to describe the dynamics of the
coordinates of the other factor, since if $\Psi=\psi(q_k)\cdot\phi(...)$,
one has $g_k^{\Psi}(t)=g_k^{\psi}(t)$.}.

Concluding, the appearance of integrated expressions has physically
unacceptable implications. In the next Section we will show how to get rid
of this bad feature of the just considered model.

\section{A more palatable model}
As everybody knows, the curl of a pseudovector transforms like a vector; this
suggests to consider, e.g., the following choice for the additional velocity
field, which is directly obtained from the expression\footnote{A completely
analogous trick can obviously be used with reference to the
expression (8.1).} (8.3):
\begin{equation}\label{10.1}
\bf v_{\it Ak}(\bf q;\it t)\propto\frac{\nabla_{\it
k}\times(\nabla_{\it k}|\rm\Psi(\bf q ;\it t)|^{\rm 2}\times\nabla_{\it
k}^{\rm 2}\bf
v_{\it Bk}(\bf q ;\it t))}{\rm|\Psi(\bf q;\it t)|^{\rm 2}}.
\end{equation}
Analogously, one could use the same trick on the expression (\ref{8.6}):
\begin{equation}\label{10.2}
\bf v_{\it Ak}(\bf q;\it t)\propto\frac{\nabla_{\it k}\times\left\{\nabla_{\it
k}\times\left \{\rm|\Psi|^{\rm 2}\left [\bf v_{\it Bk}-\sum_{\it j}\left
(\frac{\it w_{\it kj}}{\sum_{\it r}\it w_{\it kr}}\right ) \bf v_{\it
Bj}\right ]\right \}\right\}}{\rm|\Psi|^{\rm 2}}.
\end{equation}

The two just considered forms (\ref{10.1}) and (\ref{10.2}) for the
additional velocity field do not exhibit any of the unacceptable
features we have analysed in the previous Sections. More specifically, they
have the correct transformation properties under the full Galilei group,
they guarantee that the dynamics of a group of particles which are
non-interacting and not entangled with other particles is independent of
the last, and they consent to disregard the empty wave in the
case of far away superpositions of states associated to a
macroscopic object. Obviously, what we have presented are only two of many
other possible choices we will not consider since our purpose is not to
identify the most general bohmian-like model, but only to make clear that
different alternatives are possible.

\section{The case of particles with spin}
If one limits one's considerations to particles with spin, then the
elaboration of alternative models to Bohmian Mechanics becomes even
easier. Actually Bohm and Hiley themselves \cite{BH} entertained
the idea of theories exhibiting trajectories differing from the
standard bohmian ones but reproducing the quantum
configuration density distributions in the case of a particle of spin 1/2. The
procedure is quite simple, it consists in adding to the velocity field:
\begin{equation}\label{11.1}
\bf v_{\it B}(q;\it t)=\frac{\hbar}{\rm 2\it mi}\frac
{\rm\Psi^{\dagger}(\bf q;\it t)\nabla\rm\Psi(\bf q;\it t)-
[\nabla\rm\Psi^{\dagger}(\bf q;\it t)]\rm\Psi(\bf q;\it
t)}{\rm\Psi^{\dagger}(\bf q;\it t)\rm\Psi(\bf q;\it t)},
\end{equation}
proposed by Bohm and by Bell \cite{JBell} as a straightforward
generalisation of the bohmian prescription, an additional velocity field
having the following form:
\begin{equation}\label{11.2}
\bf v_{\it A}(\bf q;\it t)\propto\frac
{\nabla\times(\rm\Psi^{\dagger}(\bf q;\it t)\mbox{\boldmath$\sigma$\unboldmath}
\rm\Psi(\bf q;\it t))}{\rm\Psi^{\dagger}(\bf q;\it t)\rm\Psi(\bf q;\it t)}.
\end{equation}
The resulting velocity field
\begin{equation}\label{11.3}
\bf v_{\it N}(\bf q;\it t)=\bf v_{\it B}(\bf q;\it t)+\bf v_{\it A}(\bf q;\it t)
\end{equation}
is easily proved to have all properties which guarantee the
continuity and regularity of the trajectories, as well as all
desired properties to ensure the necessary covariance properties of
the resulting theory. The choices (\ref{11.2}) and (\ref{11.3}) represent
only one of many others which one can easily make by taking advantage,
e.g., of some of the
proposals we have presented in the previous Sections.

Just to exhibit explicit examples, let us consider all particles with spin
1/2 of our system and the
associated velocity fields $\bf v_{\it Bj}(q;\it t)$ of Standard Bohmian
Mechanics:
\begin{equation}\label{11.4}
\bf v_{\it Bj}(\bf q;\it t)=\frac{\hbar}{\rm 2\it mi}\frac
{\rm\Psi^{\dagger}(\bf q;\it t)\nabla_{\it j}\rm\Psi(\bf q;\it
t)-[\nabla_{\it j}\rm\Psi^{\dagger}(\bf q;\it t)]\rm\Psi(\bf q;\it
t)}{\rm\Psi^{\dagger}(\bf q;\it t)\rm\Psi(\bf q;\it t)}.
\end{equation}

In terms of these velocities we define scalar quantities $w_{kj}$ ($k$
and $j$ running over all particles of spin 1/2) according to Eq.\
(\ref{8.5}). Such quantities vanish whenever the considered particles are
not entangled with each other. For all spin 1/2 fermions of our
system we add to the velocity field $\bf v_{\it Bj}(\bf q;\it t)$ an
additional velocity field $\bf v_{\it Ak}(\bf q;\it t)$
according to:
\begin{equation}\label{11.5}
\bf v_{\it Ak}(\bf q;\it t)\propto\frac{\nabla_{\it
k}\times(\rm\Psi^{\dagger}(\bf q;\it t)\sum_{\it j}\it w_{\it
kj}\;\mbox{\boldmath$\sigma$\unboldmath}^{(\it
k)}\times\mbox{\boldmath$\sigma$\unboldmath}^{(\it
j)}\rm\Psi(\bf q;\it t))}{\rm\Psi^{\dagger}(\bf q;\it t)\rm\Psi(\bf q;\it t)}.
\end{equation}
In the above equation $\mbox{\boldmath$\sigma$\unboldmath}^{(\it k)}$ and
$\mbox{\boldmath$\sigma$\unboldmath}^{(\it j)}$ are the vector spin operators
of particles $k$ and $j$ respectively.

As one can immediately see, the additional velocities satisfy all necessary
requirements concerning their transformation properties for the extended
Galilei group. The additional velocity for a fermion depends only on the
positions of the fermions with which it is entangled. Furthermore the
model solves the macro-objectification problem just as Bohmian Mechanics does.

The conclusion should be obvious: the model we have just introduced has
exactly all the same features of the corresponding Standard Bohmian model
implying however different trajectories for all particles of spin 1/2.
It goes without saying that the same procedure can be used to modify the
trajectories of all particle having spin since the transformation
properties of the additional velocity do not depend on the value of the
spin.

\section{The problem of the nodes of the wavefunction}
The models introduced in Sections 10 and 11 are acceptable candidates for
alternative bohmian-like theories equivalent to Quantum Mechanics. The only
feature which could give rise to problems derives from the fact that since
we know \cite{BDG} that the bohmian velocity field can exhibit divergences
in the set $\cal N$ of the nodes of $\Psi$ and since the alternative
models contain derivatives of such fields, they could give rise to
physically unacceptable consequences. We do not intend to analyse this point
in detail. However we remark that under the conditions imposed to the
potential and to the initial wavefunction in \cite{BDG} the authors have
been able to prove that the bohmian velocity field $\bf v_{\it Bk}(q;\it
t)$ at any time is of class $\rm C^{\infty}$ on the complement of $\cal N$.
This in turn implies
that the functions $\nabla_{\it k}^2\bf v_{\it Bk}(q;\it t)$ and $w_{kj}$
appearing
in (10.1), (10.2) and (11.5) can be singular only on $\cal N$. Since the
function\footnote{Note that this function is positive and bounded by 1, is
invariant under the extended Galilei group and has the good
property underlined in the previous Sections concerning the factorizability
of the wavefunction. Actually, to be rigorous, one should take into account
that, as S.~Goldstein suggested us, in very pathological cases as for
example a wavefunction having an accumulation point X of nodes, the factor
we propose to insert could have no limit for its argument tending to X.
However, when similar situations are taken into account, problems can arise
also in Bohmian Mechanics.}
$\exp\left[-\left(\frac{\hbar}{m_kc}\frac{\nabla_k|\Psi|^2}{|\Psi|^2}\right)
^{\rm 2 }\right]$
vanishes exponentially at the nodes of $\Psi$, if one inserts such a
function as a
factor in the argument of the curl appearing in the above equations, one
gets rid of any singular behaviour of the additional velocity on the set
$\cal N$. Concluding, we have shown that there actually exist perfectly
acceptable (even though formally less simple and elegant) alternatives to
Bohmian Mechanics.

\section{The guidance view versus the Newtonian view of Bohmian Mechanics}
In this Section we take into account some recent remarks about the
interpretation of Bohmian Mechanics, which have been presented by
Baublitz and Shimony \cite{BS} and which have been reiterated
by Shimony himself in private correspondence on the content of the
present paper. The position of these authors can be
summarized in the following terms:
\begin{list}{-}{}
\item There are two possible attitudes towards Bohmian Mechanics,
 the Newtonian one, in
which one considers a classical equation of motion involving the
acceleration of the particle(s) and adds the quantum potential to
the classical one. According to the authors, the
main advantage of this approach derives from its being more
classical. However --- they also point out --- within a Newtonian
picture one should consider the initial velocities of the particles
as contingencies, so that the request that they satisfy the initial
guidance condition
\begin{equation}\label{12.1}
\bf p(\rm 0)=\nabla\it S(\bf r;\rm 0)
\end{equation}
($S/\hbar$ being the initial phase of the wavefunction) does not fit
within such a picture. In spite of this, the authors of \cite{BS} seem
to prefer this view with respect to the position which considers
only the general guidance equation we have presented in Section 1
and which is adopted by modern supporters of Bohmian Mechanics.
Baublitz and Shimony hope, with Bohm, that one could get rid of the
condition (\ref{12.1}) by adding nonlinear terms to the evolution
equation which should lead to the satisfaction of the guidance
condition in a short time, independently of the initial choice for
the velocities. In private correspondence Shimony has raised the
question of whether a similar program could be
developed for the
\it alternative
\rm bohmian models we are considering here.
\item The same authors analyse the guidance version of Bohmian Mechanics
and they recognise that, in a sense, it is more consistent than the
Newtonian one. However --- they stress --- such an
approach is less palatable because, if classically interpreted, it
seems to imply that the force on a system determines its velocity
and not its acceleration. From a philosophical standpoint they see
the guidance view as more Aristotelian and as such less
satisfactory than the Newtonian interpretation.
\end{list}

Our personal opinion on this delicate matter is quite different. Actually
we consider as misleading the
attempts to present Bohmian Mechanics as an \it almost classical
\rm theory. We agree perfectly with Bell \cite{Bll}
that it is a scandal that Bohmian Mechanics is not presented in
courses of Quantum Mechanics and precisely for the reason that its
consideration helps the student to clearly grasp the dramatic
divergences between the classical and the quantum views of natural
phenomena. Why should one try to maintain that, after all, Bohmian
Mechanics represents some sort of integration of quantum phenomena
in a classical picture when we all know very well that its most
fundamental and essential features are nonclassical? Does not such
a suggestion mislead the student making it more difficult for him to
fully appreciate the peculiar nonclassical aspects of nonlocality,
contextuality and of the unavoidable invasivity of any measurement,
which characterise the bohmian picture of physical processes?

These statements should have made clear why we adhere to the
guidance interpretation of the formalism. We would like to add some
comments. The proposal of a modified theory in which the guidance
condition is not imposed but emerges as a consequence of some
specific dynamical feature can be usefully contemplated but, at the
present stage, it represents more wishful thinking than a serious
scientific attempt. Furthermore, the problem we want to tackle in
this paper should be very clear: are there \it alternative
\rm theories to Bohmian Mechanics attributing, as it does,
\it deterministic \rm trajectories to the particles and which turn out to
be equivalent to SQM? We have deliberately chosen not to consider
the possibility of theories which, in one way or another, contradict
quantum predictions. In fact, in our opinion, if one chooses to
contemplate such a possibility it is more interesting to follow the
line of thought of the dynamical reduction program \cite{GR} than the one of
the incompleteness of the Hilbert space description
of individual physical systems.

Having stated that, we recognise that the questions raised by
Shimony of whether one can consider the Newtonian version of the
alternative models we are envisaging and whether, within such a  version,
one could consider nonlinear modifications leading to the dynamical
 satisfaction of the guidance condition are surely interesting.
Concerning the first point the answer is clearly affirmative, as
everybody would easily guess. We have presented in Appendix B the
Hamilton--Jacobi version of our alternative model. From this point
of view one can also consider the whole family of equivalent
bohmian-like theories with different trajectories as a family of
Newtonian models of Quantum Mechanics. Concerning the second point
we call attention to the fact that, at the moment, there are no indications
that it can be consistently followed even for the standard
bohmian version of the theory; therefore we consider extraneous to
the purpose of this paper to tackle this problem. It could be an
interesting task for people committed to the Newtonian
interpretation of the model. However, we cannot avoid stressing
that we do not see any reason whatsoever why, in case the indicated
line would turn out to be viable, it would be so for Bohmian
Mechanics and not for its modifications we have considered here.

\section{The Conceptual Implications of the Previous Results}
In this Section we will analyse the implications of the fact that there
exist infinitely many deterministic theories (even if they can look much
more exotic and/or less elegant) which account for physical processes in
terms of particles which follow precise trajectories, all of which
reproduce the configuration density distributions of SQM, satisfy all
possible physically meaningful requirements one can put forward for them
and nevertheless attribute different trajectories to the particles.

To put the attempts we have presented in this paper in the appropriate
light, it is useful to make precise the general perspective which we adopt.
We are interested in theories which attribute a privileged role to the
positions of the particles. Thus, the common feature of all the theories we
will analyse will be that of yielding a configuration density distribution
coinciding, at any time, with the one of SQM, i.e.\ $|\Psi({\bf q};t)|^2$,
$\bf q$ being as usual a shorthand for the position variables of all the
particles of the universe.

Obviously, within the considered class of possible theoretical models, one
should make precise in which sense they must yield the desired result. From
this point of view we can consider three different general approaches:
\newcounter{list}
\begin{list}{\arabic{list}. }%
{\usecounter{list}}
\item Deterministic models, i.e.\ models \`{a} la Bohm, in which the
particles follow precise trajectories uniquely determined by the initial
positions. The stochastic features of
such models derive simply from the lack of knowledge that one has about the
precise initial positions of the particles.
\item Stochastic models [17,18] which are based on a simple idea, i.e.\
that particles have positions and follow trajectories, but contain
irreducible (i.e.\ non-epistemic) random elements. Typically, within such
an approach, the knowledge of the wavefunction and the initial position of
a particle allows only to make probabilistic assertions about its future
position.
\item The extreme case we can consider is the one in which one
claims that all particles have definite positions at all times but he does
not commit himself, or even he denies the possibility of considering
trajectories, i.e.\ of connecting the present position of a particle to its
past position. The possibility of taking such a point of view has been
stressed by Bell \cite{JB3} in connection with a particular attitude about
the Many
World Interpretation of SQM, according to which {\it one can redistribute
the configuration {\bf q} at random (with weight $|\Psi({\bf q};t)|^2$)
from one instant to the next}. Such a position can be consistently taken
for, as stated by Bell himself, {\it we have not access to the past, but
only to memories, and these memories are just part of the instantaneous
configuration of the world}.
\end{list}

With reference to the three positions just outlined we stress the following
points. If one takes position 3 one simply denies the possibility of
describing the world in terms of particle trajectories. A detailed
discussion of position 2 has been given in ref.\ \cite{DV}. For what
concerns the problem we are analysing in this paper we note that Goldstein,
D\"{u}rr  and Zangh\`{\i} have
proved \cite{DGZ2} that in Stochastic Mechanics one can exhibit infinitely many
equivalent theories (again in the sense that they give rise to the same
distribution $|\Psi({\bf q};t)|^2)$ which are characterised by the fact
that the families of possible trajectories differ. To give an idea of the
procedure we recall that Nelson's approach identifies the quantum evolution
with a classical diffusion process of the Wiener type characterised by a
drift $\mu({\bf q};t)$ (which in turn is determined by the initial
wavefunction $\Psi({\bf q};t_0)$) and variances $\sigma_k^2$ related to the
Planck constant and to the mass of the various particles according to
$\displaystyle{\sigma_k^2=\frac{\hbar}{m_k}}$. The above mentioned authors
have proved that one can change the variance and the drift, getting however
the same density distribution. In particular, in the limit
$\sigma_k^2\rightarrow 0$ the resulting theory has deterministic
trajectories and actually coincides with Bohmian Mechanics; in the limit
$\sigma_k^2\rightarrow \infty$ the trajectories lose any direct physical
significance so that one is left only with the density distribution
$|\Psi({\bf q};t)|^2$.

The analysis of the present paper shows that a similar situation occurs in
the case of theories with deterministic trajectories, i.e.\ that Bohmian
Mechanics is a member of an equivalence class of theories, all reproducing
the desired density distribution $|\Psi({\bf q};t)|^2)$. Physically
motivated conditions, such as the genuine covariance requirement and the
others we have analysed in this paper, restrict remarkably the class of
deterministic models of the considered type but they do not allow to pick
up one member of the equivalence class as physically preferable to
all the others (obviously one can say that Bohmian Mechanics is more simple
and more elegant but we do not want to rely on such criteria). This being
the situation, we think it appropriate to conclude with some general
remarks.

We are not at all worried by the fact that one can have empirically
equivalent but essentially different theories. Actually this is precisely
what happens
with Bohmian Mechanics and Quantum Mechanics. An analogous situation
occurs in relation with the dynamical reduction models. In fact, one can
certainly build different theories which are all consistent with all known
facts about the behaviour of microsystems and which lead to the
macro-objectification of macroscopic properties in quite similar ways. The
best explicit examples are the discrete \cite{GR} (GRW) and continuous
\cite{PGP} (CSL) versions of spontaneous localisation models or the
original CSL model \cite{PGP} and those relating reduction to gravity or
mass density \cite{GGR}.

In spite of these facts we cannot avoid calling attention on some relevant
differences between the just mentioned cases (and many other similar ones
which could be mentioned) and the one we have outlined in this paper. For
instance, concerning the relations between Quantum Mechanics and Bohmian
Mechanics,
even though it is true that they turn out to be empirically equivalent, the
ontologies behind them are extremely different, such differences
representing the main point of interest of Bohmian Mechanics since they
lead to two completely different views about natural phenomena. In
particular, Bohmian Mechanics allows one to consider as fundamentally
epistemic the quantum probabilities, while SQM insists on their
nonepistemic nature. Bohmian Mechanics allows a unified description of all
natural processes while SQM suffers of an unavoidable ambiguity about the
facts of our experience. Coming to the dynamical reduction models these
qualify themselves as rival theories to SQM rather than reinterpretations
of it, and can, in principle, be experimentally tested against Quantum
Mechanics itself. Moreover the above mentioned many variants of the
dynamical reduction theories have, actually, different physical
implications, and, in any case, embody physically different mechanisms (in
spite of the fact that it can turn out to be extremely difficult to
discriminate among them).

The situation is quite different for the case we have considered in this
paper. We are confronted with alternative theories which are empirically
equivalent and have exactly the same ontology: to use Bell9s words what
these theories are about, what they speak of, is the fundamental fact that
particles \it have  definite positions at all times \rm and \it move along
precise trajectories \rm (which, however, as we all know, cannot be
detected).

We can then imagine various possible scenarios:
\begin{description}
\item [ ({\it a}) ] Our hope is that someone (more expert than we are on
this topic) could
find physically meaningful and compelling criteria to identify, among the
class of theories \it equivalent \rm to Bohmian Mechanics in the sense we
have made precise, one which is physically privileged. One could, e.g.,
take the argument of Bohm and Hiley as implying that the nonrelativistic
limit of Dirac
equation seems to favour a non-standard bohmian trajectory with
respect to the standard one. Such a remark could be used to
identify criteria to pick up the best model among all possible
ones. If this would be feasible, then one would have surely
achieved a deeper understanding of the theory.
\item [ ({\it b}) ] Alternatively, it could happen that, at the
nonrelativistic level
characterising the present analysis, no compelling criterion (besides
formal simplicity) exists making one of the equivalent versions physically
better grounded than the others. In this case one would be led to the
conclusion that, actually, the appropriate bohmian-like reformulation of
Quantum Mechanics is given by the equivalence class [$\cal B$] of all
theories which attribute definite (but different) trajectories to
the particles, reproducing the probability density distribution of
SQM. This fact would not be distressing and could turn out to be only
temporary and of some use. In fact, it could very well happen that the
hypothetical relativistic generalisations of the theory would not be
affected by the above ambiguity. One could even hope that the greater
variety of nonrelativistic bohmian-like theories discussed in this paper
could help in finding a satisfactory relativistic generalisation. Having
more freedom in the theory one is dealing with usually is an advantage and
may allow to circumvent more easily some difficulties.
\item [ ({\it c}) ] For completeness we cannot avoid mentioning that one,
taking a
radical position about our analysis, could consider the fact that many
inequivalent theories with different trajectories can be exhibited as an
indication that one must not consider the trajectories since they are ill
defined. In other words one could be led to take a position like the one we
have mentioned above as position 3.

We do not share such an attitude and we stress that, in the last issue, it
would correspond to taking a quite trivial and ambiguous perspective about
the conceptual problems of Quantum Mechanics. The best illustration of this
fact is given once more by a sentence by Bell [19]: {\it Everett's
replacement of the past by memories is a radical solipsism --- extending to
the temporal dimension the replacement of everything outside my head by my
impressions, of ordinary solipsism or positivism. Solipsism cannot be
refuted. But if such a theory were taken seriously it would hardly be
possible to take anything else seriously. ... It is always interesting to
find that solipsists and positivists, when they have children, have life
insurance}. This shows that the possibility just envisaged would turn out
to be almost devoid of any physical significance
and would practically amount to an \it a priori \rm refusal of the bohmian
program.
\end{description}
\vspace{0.5in}
\noindent\bf Acknowledgements: \rm we acknowledge many discussions with
D.~D\"{u}rr and  N.~Zangh\`{\i}, whose remarks have played an extremely
important role for this paper. We also
thank G.~Bacciagaluppi, G.~Calucci, S.~Goldstein, P.~Pearle,
A.~Rimini and A.~Shimony for stimulating remarks.
\vspace{0.5in}
\pagebreak

\begin{center}
\LARGE\bf Appendix
\end{center}
\vspace{0.6in}
\appendix
\section{Gauge Transformations}
Suppose we consider the Hamiltonian $\displaystyle{\hat
H=\frac{\bf\hat p^{\rm 2}}{\rm 2\it m}+V(\bf\hat r)}$ and let us take
into account the solution $|\Psi;t\rangle$ of the Schr\"{o}dinger
equation:
\begin{equation}\label{A.2.1}
i\hbar\frac{\partial |\Psi,t\rangle}{\partial t}=\hat
H|\Psi;t\rangle;
\end{equation}
then we know that for any real function $\alpha(\bf\hat r)$:
\begin{equation}\label{A.2.2}
|\tilde\Psi;t\rangle=e^{i\alpha(\bf\hat r)}|\Psi;t\rangle
\end{equation}
is a solution of the Schr\"{o}dinger equation for the Hamiltonian:
\begin{equation}\label{A.2.3}
\tilde H=e^{i\alpha(\bf\hat r)}\hat H e^{-i\alpha(\bf\hat r)}=
\frac{[\bf\hat p-\it\hbar\nabla\alpha\bf(\hat r)]^{\rm2}}{\rm 2\it
m}+V(\bf\hat r),
\end{equation}
i.e.:
\begin{equation}\label{A.2.4}
i\hbar\frac{\partial |\tilde\Psi,t\rangle}{\partial t}=\tilde H
|\tilde\Psi;t\rangle.
\end{equation}
Note that, since the coordinate representations of $|\Psi,t\rangle$
and $|\tilde\Psi,t\rangle$ are $\Psi(\bf r,\it t)$ and
$\tilde\Psi(\bf r,\it t)=e^{i\alpha(\bf\hat r)}\rm\Psi(\bf r,\it
t)$ respectively, one has $|\tilde\Psi(\bf r,\it
t)|^{\rm2}=|\rm\Psi(\bf r,\it t)|^{\rm2}$ $\forall (\bf r;\it t)$,
and the configuration density distribution is then the same at all
times and in all places for the two physical systems under
examination.

Let us now evaluate the velocity field associated to
$\tilde\Psi(\bf r;\it t)$ by using the standard expression of
Bohmian Mechanics associated to Eq.\ (A.1):
\begin{equation}\label{A.2.5}
\bf\tilde v_{\it B}(r;\it t)=\frac{\hbar}{m}\Im\frac{\rm\tilde\Psi^{\ast}
(\bf r;\it t)\nabla\rm\tilde\Psi(\bf r,\it t)}{|\rm\tilde\Psi(\bf
r;\it t)|^{\rm 2}};
\end{equation}
we then have:
\begin{equation}\label{A.2.6}
\bf\tilde v_{\it B}(r;\it t)=\frac{\hbar}{m}\Im\frac{\rm\Psi^{\ast}
(\bf r,\it t)\nabla\rm\Psi(\bf r,\it t)}{|\rm\Psi(\bf r,\it
t)|^{\rm 2}}+\frac{\hbar}{\it m}\nabla\alpha(\bf r).
\end{equation}
However, it has to be mentioned that when the Hamiltonian contains
momentum terms of the type we have introduced, the conserved
current density for the Schr\"{o}dinger equation has to be changed
(as one can easily check) according to:
\begin{equation}\label{A.2.7}
\bf j(r;\it t)\longrightarrow\bf j(r;\it t)-\frac{\hbar}{m}
|\rm\Psi(\bf r,\it t)|^{\rm 2}\nabla\alpha(\bf r).
\end{equation}
Since the velocity is given by the ratio of the current and the
density, we see then that the appropriate velocity field for the
wavefunction $\tilde\Psi(\bf r,\it t)$ is:
\begin{eqnarray}
\bf\tilde v_{\it B}(\bf r,\it t)&=&\frac{\hbar}{m}\Im\frac{\tilde\Psi^{\ast}
(\bf r,\it t)\nabla\rm\tilde\Psi(\bf r,\it t)}{|\rm\Psi(\bf r,\it
t)|^{\rm 2}}-\frac{\hbar}{\it m}\nabla\alpha(\bf r)\nonumber\\
&=&\frac{\hbar}{m}\Im\frac{\Psi^{\ast} (\bf r,\it
t)\nabla\rm\Psi(\bf r,\it t)}{|\rm\Psi(\bf r,\it t)|^{\rm
2}}+\frac{\hbar}{\it m}\nabla\alpha(\bf r)-\frac{\it\hbar} {\it
m}\nabla\alpha(\bf r)\\ &=&\frac{\hbar}{m}\Im\frac{\Psi^{\ast} (\bf
r,\it t)\nabla\rm\Psi(\bf r,\it t)}{|\rm\Psi(\bf r,\it t)|^{\rm
2}}=\bf v_{\it B}(\bf r,\it t),\nonumber
\end{eqnarray}
as it had to be, since the transformation we have considered
corresponds simply to changing the phases of the positions
eigenstates, leaving the physics unchanged.

Thus, even though a gauge transformation implies a change in the
wavefunction (but not in the associated probability density),
within a bohmian picture it does not change the trajectories of the
particles, an important fact for the alternatives to the bohmian
theory we are envisaging in this paper.

\section{Hamiltonian Formulation of the Theory}
In this appendix we will show how it is possible to give a
hamiltonian formulation to our alternative theory, following the
same procedure pursued by Bohm \cite{DB1} in 1952.

First of all let us consider for simplicity a system of one
particle and take the Bohm point of view writing $\Psi$ as
\begin{equation}\label{1}
\Psi(\bf r;\it t)=R(\bf r;\it t)\exp\left[\frac{i}{\hbar}S(\bf r;\it t)\right].
\end{equation}
Then the Schr\"{o}dinger equation is equivalent to:
\begin{eqnarray}
&&\spc\frac{\partial S}{\partial t}+\frac{(\nabla S)^2}{2m}+V+Q=0
\\ &&\spc\frac{\partial R^2}{\partial t}+\nabla\cdot
(R^2\frac{\nabla S}{m})=0\label{3}
\end{eqnarray}
where $V$ and $\displaystyle{Q=-\frac{\hbar
^2}{2m}\frac{\nabla^2 R}{R}}$ are the actual and the quantum
potentials, respectively.

Since the bohmian momentum is given by:
\begin{equation}\label{4}
\bf p=\it m\bf v=\nabla\it S
\end{equation}
we obtain that the motion implied by Eq.\ (\ref{4}) is equal to
that governed by the Hamiltonian of the form
$\displaystyle{\hat{H}=\frac{\hat{p}^2}{2m}+V+Q}$.

Taking the same point of view, we note that Eq.\ (\ref{1}) can also
be rewritten as:
\begin{equation}\label{5}
\frac{\partial S}{\partial t}+\frac{(\nabla S+m\bf v_{\it A})^{\rm 2}}{\rm2\it
m}+V+\tilde Q=0
\end{equation}
with $\displaystyle{\tilde Q=Q-\frac{m}{2}\bf v_{\it A}^{\rm2}-\nabla\it
S\cdot\bf v_{\it A}}$. It is then clear that Eq.\
(\ref{5}) can be considered as the Hamilton-Jacobi equation for a
particle moving in a velocity dependent generalised potential, like
for example a particle of charge $-e$ in a magnetic field with a
vector potential $\displaystyle{\bf A=-\it
\frac{mc}{e}\bf v_{\it A}}$ and a scalar potential $\tilde{V}=~V+\tilde Q$.
Therefore we have shown that also our trajectories admit a
Hamiltonian description. We conclude by proving that the
generalised potential relative to the system described by Eq.\
(\ref{5}) is given by
\begin{equation}\label{6}
\tilde{V}_{\mbox{\scriptsize gen}}=\tilde{V}+m\bf v\cdot v_{\it A}.
\end{equation}
In fact, taking into account Eq.\ (\ref{5}) (more precisely the
equation obtained by taking the gradient of the two sides of Eq.\
(\ref{5})), it is simple to check that:
\begin{eqnarray}\label{7}
m\frac{dv_{Nx}}{\it dt}&=&-\frac{\partial\tilde{V}_{\mbox{\scriptsize
gen}}}{\partial x}+\frac{d}{dt}
\frac{\partial\tilde{V}_{\mbox{\scriptsize gen}}}{\partial
v_x}\nonumber\\ &=&-\frac{\partial\tilde{V}}{\partial x}-m\bf
v\cdot
\frac{\partial\bf v_{\it A}}{\partial\it x}+\it m\frac{dv_{\it Ax}}{\it
dt}\nonumber\\
&=&\frac{\partial}{\partial t}(\partial_x S)+\frac{\nabla S}{\it m}
\cdot\nabla(\partial_x S)+\bf v_{\it A}\cdot\it\partial_x(\nabla\it S)+\it
m\frac{dv_{\it Ax}}{\it dt}\\
&=&\frac{\partial}{\partial t}(\partial_x S)+\frac{\nabla S+m\bf v_{\it
A}}{\it m}
\cdot\nabla(\partial_x S)+\it m\frac{dv_{\it Ax}}{\it dt}\nonumber\\
&=&\frac{d}{dt}(\partial_x S)+\it m\frac{dv_{\it Ax}}{\it dt}\nonumber\\
&=&\it m\frac{d}{dt}(v_{Bx}+ v_{Ax}).
\end{eqnarray}

\end{document}